\documentclass[a4paper]{spie}  

\usepackage{graphicx}
\usepackage{scalerel}
\usepackage{comment}
\usepackage{booktabs}
\usepackage{tabularx}
\usepackage{tensor}

\usepackage{subcaption}
\usepackage{multirow}
\usepackage{natbib} 
\usepackage{amsmath,amsfonts,amssymb}
\usepackage{euclid}

\usepackage[colorlinks=true, allcolors=blue]{hyperref}

\usepackage[table]{xcolor}

\makeatletter
\def\@maketitle{\newpage
 \null
 \if0\@ptsize\vspace{-10mm}\else\vspace{-12mm}\fi
 \begin{center}%
  \ifcmr
    \if0\@ptsize {\typeout{10pt}\setlength{\baselineskip}{19.2pt} \@title \par}\fi
    \if1\@ptsize {\typeout{11pt}\setlength{\baselineskip}{20.4pt} \@title \par}\fi
    \if2\@ptsize {\typeout{12pt}\setlength{\baselineskip}{23pt} \@title \par}\fi
  \else
    {\@title \par}
  \fi
  \vskip 3.5mm
  {\large \@author \par}   
  \vskip 1.5ex
  {\large \@date}%
 \end{center}%
 \par
}

\def\maketitle{\par
 \settitlefont
 \setamplefont
 \begingroup
   \def\thefootnote{\fnsymbol{footnote}}%
   \def\@makefnmark{\hbox to\z@{$\m@th^{\@thefnmark}$\hss}}%
   \if@twocolumn
     \twocolumn[\@maketitle]%
   \else
     \newpage
     \global\@topnum\z@
     \@maketitle
   \fi
   \@thanks
 \endgroup
 \let\maketitle\relax
 \let\@maketitle\relax
 \gdef\@thanks{}\gdef\@author{}\gdef\@title{}\let\thanks\relax
}
\makeatother

\title{Characterisation Methodology of Persistence and Anti-Persistence in IR Detectors for the Euclid and SVOM Space Missions\thanks{\hspace{1mm} This paper is published on behalf of the Euclid Consortium}}

\author[a]{Naomie De Araujo }
\author[a]{Aurélia Secroun}
\author[a]{Marie Tourneur Silvain}
\author[a]{Éric Kajfasz}
\author[a]{Eliott Robineau}
\author[a]{Alioune Diop}
\affil[a]{Centre de Physique des Particules de Marseille, Marseille, France}

\authorinfo{Further author information: (Send correspondence to N. De Araujo)\\
N. De Araujo: E-mail: \href{mailto:dearaujo@cppm.in2p3.fr}{dearaujo@cppm.in2p3.fr}\\
A. Secroun: E-mail: \href{mailto:aurelia.secroun@cppm.in2p3.fr}{secroun@cppm.in2p3.fr}}

\begin{document} 
\pagestyle{plain} 
\maketitle

\begin{abstract}

H2RG and ALFA are the infrared detectors used in the \textit{Euclid} and SVOM space missions. These types of detectors may be affected by persistence and anti-persistence, phenomena that distort pixel responses after exposure to light due to simultaneous trapping and release of charges within the pixel. These phenomena have shown to cause biases in flux measurements and potentially false detections, among other effects. This article presents a universal methodology for characterising persistence and anti-persistence using differential ramps, Savitzky--Golay filtering, and non-linearity correction. When applied to ground characterisation data, this approach reveals spatial inhomogeneities, environmental dependencies (in particular fluence, flux, temperature and wavelength), and unexpected spectral effects in ALFA. These preliminary results validate the method and pave the way for in-flight correction models capable of meeting the missions' accuracy requirements.

\end{abstract}


\tableofcontents
   
\section{\label{sc:Intro}Introduction}

The \textit{Euclid} mission aims to create a 3D cartography of the Universe to uncover the nature of the dark sector, specifically dark energy and dark matter \citep{Euclid2025a}. To achieve this, it is equipped with two main instruments: the VIS camera, which measures weak gravitational lensing, and the NISP (Near-Infrared Spectrometer and Photometer). NISP features a large focal plane array of 16 H2RG MCT (mercury-cadmium-telluride) detectors and performs near-infrared photometry and slit-less spectroscopy, enabling the measurement of galaxy redshift and the mapping of their 3D distribution \citep{Euclid2025b}. By combining these observations, \textit{Euclid} will map the large-scale structure of the Universe, trace its expansion history, and constrain the properties of dark energy and dark matter over cosmic time.

The SVOM mission aims to detect and study gamma-ray bursts (GRBs) and other high-energy transient phenomena to probe the extreme Universe. This is made possible by instruments such as the CAGIRE camera, featuring the ALFA detector, enabling wide-field observations in the near-infrared domain \citep{Fortin2024}.

In recent years, MCT CMOS infrared (IR) detectors have been increasingly used in both space missions and ground-based telescope instrumentation. However, they all face a common challenge: persistence. In this context, persistence refers to a measured signal that exceeds its expected value. It results from the release of a fraction of the signal trapped in the detector's sensitive layer during previous acquisitions. Conversely, anti-persistence corresponds to the signal loss due to trapping. Since its launch in July 2023, the \textit{Euclid} mission has observed persistence effects in its detectors. In-flight data confirm the presence of this phenomenon -- also previously detected during ground testing -- and suggest that the effect may be even more pronounced in space \cite{Conseil2026}. In the SVOM mission, the ALFA detector will observe the infrared afterglows of GRBs -- fading emissions that follow these explosive events. However, their decaying signals can be mistaken for persistence, potentially causing false detections.

Carefully accounting for these effects is essential to ensure the accuracy of cosmological measurements. They manifest as long-lasting signatures, sometimes persisting for several hours after exposure to intense light sources, and often exhibit specific temporal evolution and spatial structures across the focal plane. 

\begin{figure}[!htbp]
  \centering
  \begin{subfigure}{0.4\textwidth}
    \centering
    \includegraphics[width=\linewidth, height=0.9\linewidth]{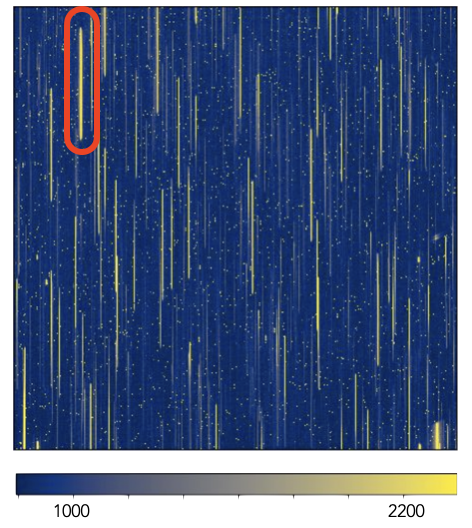}
    \vspace{0.1cm} 
    \caption{Spectroscopic acquisition}
    \label{fig:spectro}
  \end{subfigure}
  \hspace{1cm}
  \begin{subfigure}{0.4\textwidth}
    \centering
    \includegraphics[width=\linewidth, height=0.9\linewidth]{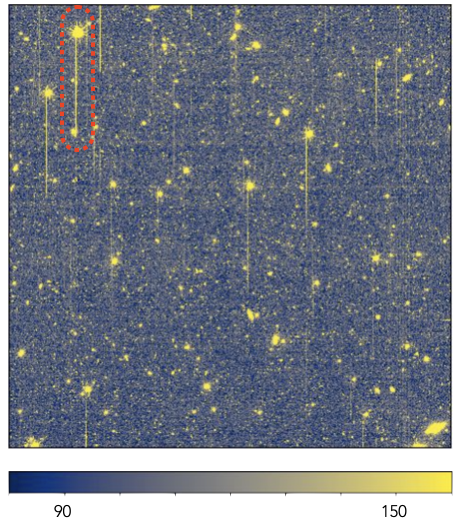}
    \vspace{0.1cm} 
    \caption{Photometric acquisition}
    \label{fig:photo}
  \end{subfigure}

  \vspace{0.4cm} 

  \caption{Extracts from a \textit{Euclid} flight H2RG acquisition taken during commissioning: (a) spectroscopic acquisition ($\simeq$ 10\,min) and (b) following photometric acquisition ($\simeq$ 1\,min) in \YE band. The persistence signal in (b) (highlighted in red) clearly lasts beyond the reset of the spectroscopic acquisition. Credits: Euclid Consortium}
  \label{fig:euclid_acquisitions}
\end{figure}

In the context of the \Euclid mission, spectroscopic acquisitions can induce persistence that remains clearly visible in subsequent photometric exposures, as illustrated in Fig. \ref{fig:euclid_acquisitions}. 
This figure shows two  acquisitions extracted from the \textit{Euclid} survey: the first is a spectroscopic measurement, characterised by distinct lines on the detector, each representing a spectrum. The second is a photometric acquisition, in which point-like features are visible, each corresponding to an observed galaxy.
However, residual lines from the spectroscopic measurement are also visible, indicating persistence effects on the detector. One example is circled in red.

Additionally, photometric observations in one band (say \YE) can also leave residual signals affecting the following bands (say \HE or \JE) in the survey sequence \cite{Laureijs2012}. Such effects introduce systematic biases that may compromise key measurements, including source identification, redshift measurements, flat-field correction, or photometric accuracy (over- or under-estimation of fluxes).

Given the strict performance requirements of the \textit{Euclid} mission -- specifically a 1\% precision in flux measurements and 95\% of operational pixels -- signal persistence poses a risk. As this phenomenon has proven more pronounced than ground-based measurements had predicted, the strategy for mitigating persistence has to be revised. Currently, the baseline approach relies on masking, with no implementation of active correction. 
The present masking alone removes between 5\% and 8\% of pixels, and this fraction may increase further during periods of high solar activity \citep{Conseil2026}.
It would thus be suitable to work with a more accurate model of persistence, describing both its temporal and spatial behaviour, as well as its environmental dependencies.

In this context, we investigate the persistence effects associated with two space missions, \textit{Euclid} and SVOM~\citep{Fortin2024}, for which the CPPM is responsible for characterising their infrared detectors. Extensive ground-based campaigns were conducted, varying flux, temperature, and wavelength to analyse the associated dependencies for both \textit{Euclid}'s H2RG and SVOM's ALFA detectors.

The challenge lies in developing a universal method for analysing infrared detector persistence that is applicable to any space mission. Our approach aims to extract the fundamental dependencies of persistence -- such as flux, temperature, and wavelength -- from ground-based datasets. To demonstrate the method's general applicability, we apply it to two distinct detectors: one of the sixteen flight H2RG from the \textit{Euclid} mission and the ALFA from the SVOM mission. By validating the method across different technologies, we show that it is independent of the specific instrument and can be adapted for other missions.
The ultimate goal is to establish a pixel response model -- incorporating (anti-)persistence -- based on ground-based data. In the long term, the aim is to integrate this model into the observation process, thereby ensuring effective mitigation of persistence effects during operations.

In light of these considerations, this paper is structured as follows: We begin by presenting the operating principle of the infrared detection technology used in the \textit{Euclid} and SVOM missions. There will be a focus on the hybrid CMOS-MCT detectors and the signal biases introduced by persistence and anti-persistence.
We then investigate the origins and mechanisms of these phenomena, drawing on the current state of the art to contextualise our contributions.
The core of the paper is dedicated to the methodology for characterising (anti-)persistence, including data pre-processing, flux measurement, and the extraction of amplitude and current parameters.
The preliminary results will demonstrate the effectiveness of this approach in quantifying dependencies on temperature, flux, and wavelength.

\section{\label{sc:Detectors} Overview of detectors}

Before detailing the technology behind hybrid detectors, let us briefly introduce the detectors under study. The H2RG detectors, developed by Teledyne (USA), equip the \textit{Euclid} mission. The ALFA detector, a unique prototype designed by Lynred (France), is integrated into the CAGIRE camera of the SVOM mission. Both represent state-of-the-art infrared detection technologies tailored to the demanding requirements of space-based observations.

\subsection{\label{sc:Detection} Structure and signal chain in hybridised IR detectors}

The two detectors under study consist of a 2048×2048 pixel CMOS sensor hybridised with a sensitive layer of MCT. This PN junction semiconductor layer acts as a photodetector, with its bandgap engineered by adjusting the mercury and cadmium composition to cover the near-IR range \citep{Rogalski2005}. \textit{Euclid}'s H2RG detectors have a cut-off wavelength of 2.3\,$\mu$m, while that of the ALFA detector of the SVOM mission is 2.5\,$\mu$m.

In such detectors, incident photons with energy greater than the bandgap generate electron-hole pairs in the semiconductor. Ideally, these photo-generated charges are collected and converted into a measurable electrical signal via the CMOS readout chain, which includes source follower transistors (SFT). The SFTs amplify the signal from each pixel before it is transmitted through the indium bumps to the CMOS electronics for further processing and digitisation.

While the H2RG and ALFA detectors are similar in design,
their sensitive layers are fabricated using distinct processes: \textit{Euclid}'s H2RGs sensitive layer is deposited in a molecular-beam epitaxy (MBE), whereas ALFA's MCT layer results from a liquid-phase epitaxy \citep{Norton2002,Gravrand2022}. This difference may have an impact on the behaviour of persistence and anti-persistence.

\subsection{\label{sc:Detection} Signal acquisition}

Both missions use non-destructive readout, allowing multiple reads of the same pixel during integration without resetting the signal. In this mode, the pixel continues to accumulate photons between successive measurements.
In practice, a full $2048\times 2048$ matrix readout is referred to as a frame, with the readout cadence and number of output channels dictating the acquisition speed. For both missions, the detectors feature 32 output channels and a readout frequency of 100 kHz. This means that, in theory, the acquisition time for a frame is 1.31 seconds. In practice, the response time of the electronics may add additional time. For the H2RG detector, a frame lasts 1.45 seconds, while for ALFA, it has not been measured. Therefore, in the study, we use the theoretical value of 1.31 seconds.

Under ideal conditions, for a homogeneous and constant incoming flux, the expected response is a linear ramp, known as an up-the-ramp (UTR) signal. 
This signal is directly proportional to the number of accumulated charges.

The ideal response, denoted $S_{\rm{ideal}} (t)$ may be simply described by Eq. \ref{eq:ideal}, expressed in ADU (for analogic-to-digital unit). $S_{e^-}(t)$ corresponds to the number of electrons collected at time $t$ and $cg$ to the conversion gain, another property of hybridised IR detectors,

\begin{equation}\label{eq:ideal} 
    S_{\rm{ideal}} (\rm{t})= \rm{cg} \times S_{e^{-}}(\rm{t}).
\end{equation}

This latter parameter defines the relationship between the digital signal (in ADU) and the physical charge (in electrons, $e^-$). It depends on multiple factors, such as the analog-to-digital converter (ADC) characteristics, and the overall sensitivity of the readout chain. However, unlike $S_{e^-}$(t), it does not depend on quantum efficiency (QE), which represents the fraction of incident photons that generate charges contributing to the signal.

Accurately determining the conversion gain is important for obtaining a usable physical quantity, ensuring unit comparability across different detectors, and enabling precise flux measurements.
However, real-world detectors are affected by intrinsic properties or defaults that introduce deviations from this ideal behaviour. Applying the standard PTC method to measure the gain without taking into account these properties induces a bias of 8\% to 9\% on this measurement, as has been shown in \cite{LeGraet2026}.

Let us mention the main properties that we consider here.
First, the dark current is a signal resulting from the thermal excitation of electrons in the sensitive layer, which generates a charge even in the absence of incident light. This current increases exponentially with temperature and constitutes a source of noise in infrared detectors. For missions like \textit{Euclid} and SVOM, detectors are cooled to cryogenic temperatures ($95\,\rm{K}$ for \textit{Euclid}’s NISP, 100\,K for SVOM's ALFA \cite{NouvelDeLaFleche2023}) to minimise its impact. Their dark current therefore does not exceed 0.01 $e^-/{\rm s}$.

A second visible defect in the signal is RON or readout noise. This is the noise added by the acquisition electronics of the detection chain, downstream of the p-n junction. For \textit{Euclid}'s H2RG detectors, $\sigma_{\rm{RON}}\simeq13\,e^-$, while for ALFA, $\sigma_{\rm{RON}}\simeq28\,e^-$. 

There is another type of readout noise: common-mode noise, which arises from temporal drifts in the output voltages of the source-follower transistors (SFTs) within a given channel or frame. To address this, hybrid CMOS detectors incorporate reference pixels along the outer edges of the array, typically within the first and last four rows and columns. Although electronically identical to active pixels, these pixels are not coupled to the HgCdTe (MCT) sensitive layer. Consequently, their signal is independent of the incident photon flux and reflects only the output voltage variations of the SFTs. This design enables the correction of common temporal drifts by frame-by-frame subtraction of the reference pixel signals, thereby effectively mitigating common-mode noise. The offsets are subtracted by the corresponding offsets of reference pixels, provided that the processing of normal and reference pixels is identical.

Another effect to introduce is inter-pixel capacitance (IPC), resulting from parasitic capacitive coupling between adjacent pixels in the detector array. When a photon-generated charge is collected in a given pixel, a fraction of the signal can `leak' into neighbouring pixels due to their proximity. Given the small pixel size -- 18$\mu$m for the H2RG and 15 $\mu$m for the ALFA --  the coupling becomes significant. This results in spatial signal spreading, with a typical IPC level of approximately 3\% for both detectors (see \cite{LeGraet2026} for measurement details and decorrelation techniques).
In the case of flat-field illumination, the IPC effect is reciprocal between adjacent pixels, leading to a compensation that effectively cancels out its spatial signature and renders it invisible in the final data. 
Other types of crosstalk are considered negligible compared to IPC and are not addressed in this analysis.

Next, there is a significant phenomenon that we expect to affect our study: non-linearities. There are several types: First, there is a non-linearity arising from the diode's non-constant response as its charge capacity fills up. This response gradually becomes less efficient, resulting in a deviation from the ideal linear response shown in Eq.~\ref{eq:ideal}.
This results in a signal decrease during constant-flux acquisition. In the literature, this response is considered to be modellable by a third- or fourth-order polynomial function \cite{Hilbert2014, Euclid2026}. This effect cannot be ignored, as it might affect anti-persistence -- a point we will address in the following sections.

Another type of non-linearity is reciprocity failure: Signals with the same fluence\footnote{The fluence is obtained by multiplying the measured flux by the duration of illumination.}, obtained from acquisitions with different flux rates, do not yield the same measured value \cite{Baril2008, Biesiadzinski2011, Tulloch2019}.
Finally, there is the saturation -- once the well capacity is filled, the detector no longer accumulates charge. Once saturated, the signal becomes constant. We will need to take into account the fluence at which the detector saturates in the proposed method.

The last effect, and not the least, is persistence and its counterpart, anti-persistence, which are at the heart of this study. Persistence refers to the residual signal remaining in a detector after exposure to a light source, whereas anti-persistence describes the signal reduction that occurs during exposure due to trapping. In Sect. \ref{sc:Persistence}, we will detail the origins and manifestations of these phenomena. Subsequently, we will describe our method for characterising persistence and anti-persistence (hereafter referred to as (anti-)persistence for brevity). Our approach accounts for the diverse properties of the detectors, aiming to effectively decouple their effects from the (anti-)persistence signals in order to isolate and precisely analyse each contribution.
\newline
\newline
Taking all these effects into account, a more realistic expression for the UTR signal in ADU, incorporating known deviations from the ideal response, is given by
\begin{equation}\label{eq:rputr}
    S_{\rm{real}}(\rm t) = (S_{\rm{ideal}}(\rm t) + S_{\rm{
trapping}}(\rm t) +S_{\rm{de-trapping}}(\rm t))\times C_{\rm{NL}} + I_{\rm{dark}} + S_{\rm{noise}}. 
\end{equation}
With  $C_{{\rm{NL}}}$ the coefficients related to non-linearity, $S_{\rm{trapping}}$ and $S_{\rm{de-trapping}}$ the signals due to trapping and de-trapping, $I_{\rm{dark}}$ the dark current, and $S_{\rm{noise}}$ the readout noise typically described by $\sigma_{{\rm{RON}}}$. This equation does not take potential correlations into account to simplify the reading.

However, direct analysis of the raw UTR signal can be complex due to the superposition of contributions from these various effects. To address this, we focus on differential ramps, obtained by subtracting two successive UTR readings for a given pixel or group of pixels. This approach offers two major advantages:

\begin{itemize}
    \item It improves the visibility and quantifiability of transient effects such as persistence and non-linearity, which often manifest as deviations from a linear trend.
    \item It allows for the subtraction of components evolving at a constant rate -- such as dark current -- during fixed-temperature acquisitions, thereby facilitating the isolation of dynamic effects.
\end{itemize}

The equation for the differential signal between frames n and n+1 is \begin{equation}\label{eq:rpdiff}
S_{\rm{diff}}(\Delta t) = S_{\rm{real}}(t_{\rm n+1}) - S_{\rm{real}}(t_{\rm n}).
\end{equation}

The resulting differential ramps reveal signatures of persistence, non-linearity, and other phenomena, as illustrated in Fig. \ref{fig:responses}. In return, this approach increases the noise by a factor of $\sqrt{2}$. The figure shows several differential flux ramps obtained at a temperature of 90\,K using the H2RG detector. Those signatures will be described  in detail in Sect. \ref{sc:flux}.

We were able to introduce several intrinsic properties of the detector -- which are essential for studying persistence -- by observing their influence on the signal during a differential ramp. This allows for a better understanding of the signal observed during differential ramps, and thus helps to isolate (anti-)persistence as effectively as possible within the proposed method.

With this understanding, we will now examine persistence and anti-persistence in detail. Knowing their definitions and mechanisms is important for establishing a characterisation method and interpreting the observed behaviours.

\subsection{\label{sc:Persistence}Description of (anti-)persistence}

In the literature, persistence is generally modelled using multiple-decay exponential functions or power-law distributions \cite{Serra2016, Legoff2022phd, Tulloch2019, Brandt2026, Fazar2025}. In the case of multiple-decay exponential functions, each exponential would represent a family or type of trap with its own characteristic de-trapping time. By contrast, anti-persistence remains largely unexplored from a theoretical perspective, and the few models proposed rely primarily on decaying exponential functions \cite{Brandt2026}.
It should be noted that the \textit{Roman} Space Telescope team uses the term `burn-in’ to describe what we define here as anti-persistence. 

In both cases, these approaches describe the phenomena through key parameters, conventionally referred to as the amplitude and the characteristic trapping/de-trapping times. Although these quantities are widely adopted, they are not the only ones proposed in the literature: other terms, such as the persistence flux, have also been used to describe persistence amplitude. More specifically, the amplitude, expressed in $e^-$, represents the number of excess charges (persistence) or charge deficits (anti-persistence) in the readout signal during a frame time. The characteristic trapping/de-trapping times expressed in $e^-/{\rm s}$, represent the timescales at which the traps release charges.

In this study, the terms amplitude (in $e^-$) and current (in $e^-/{\rm s}$) are used to describe (anti-)persistence. Here, the current represents the variation of this amplitude over time -- namely the dynamics of (anti-)persistence.

The advantage of this method lies in its ability to incorporate external dependencies, such as temperature or fluence. Subsequently, the extraction of these parameters could lead to the development of a model capable of describing the behaviour of (anti-)persistence.
An in-depth analysis of these dependencies, based on the physics of the traps, will be developed in Sect. \ref{sc:Stateoftheart}.

\subsubsection{\label{sc:def}Origin and mechanisms}

As previously mentioned, the sensitive layer of our NIR detectors is made of an MCT alloy. However, during the fabrication process, the structure is modified by defects and impurities in the crystal lattice, which introduce intermediate energy levels within the material bandgap. These defects act as traps for photo-generated charges.
Upon reset of the potential difference, the trapped charges are gradually released. The amplitudes and currents associated with trapping and de-trapping depend on environmental parameters, such as the flux, the fluence, or the temperature. They also depend on the trap density, activation energy and distribution within the sensitive layer \cite{Smith2008}.
It is also well established that persistence signals -- excess, resulting from the delayed release of trapped charges -- can persist for several hours. This behaviour is explained by trapping dynamics associated with a trap distribution characterised by varying activation energies and characteristic (de-)trapping times \citep{Smith2008,Tulloch2019,Legoff2022phd}. Similar temporal behaviour is expected for anti-persistence. 

\begin{figure}[!htbp]
  \centering
  \begin{subfigure}{0.45\textwidth}  
    \centering
    \includegraphics[height=0.7\linewidth]{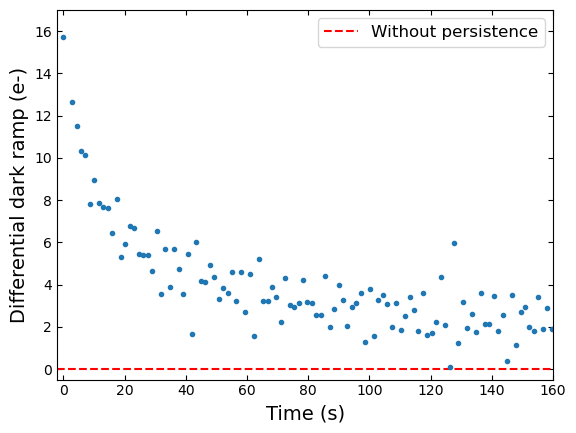}
    \caption{Differential dark ramp}
    \label{fig:persistence}
  \end{subfigure}  
  \hfill
  \begin{subfigure}{0.45\textwidth}
    \centering
    \includegraphics[height=0.7\linewidth]{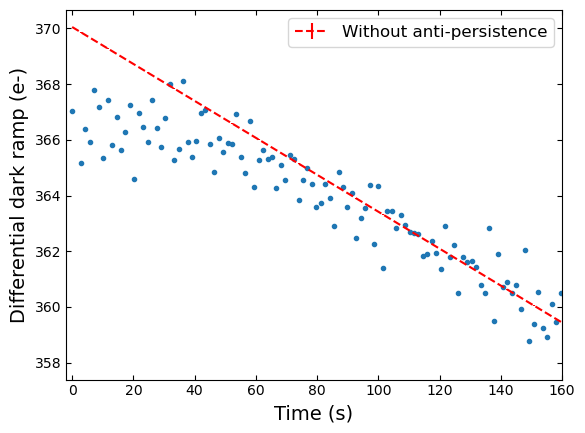}
    \caption{Differential illumination ramp}
    \label{fig:antipersistence}
  \end{subfigure}
  \vspace{0.4cm}
  \caption{H2RG detector: Highlighting differential ramp signatures of (a) persistence and (b) anti-persistence. Set-point flux = 280\,$e^-/{\rm s}$, \textit{T} = $90$\,K, signal averaged over 64$\times$64 pixels}
  \label{fig:signatures}
\end{figure}

During an acquisition, both trapping and de-trapping occur simultaneously. Depending on the conditions of  observation, one effect may dominate the other as illustrated in Fig. \ref{fig:signatures}:
\begin{itemize}
\item As seen in Fig.~\ref{fig:persistence}, in a dark differential ramp, de-trapping dominates, resulting in a signal higher than expected (blue dots), even though the theoretical signal (red dashes) is expected to be equal to 0\,$e^-/{\rm s}$ in the absence of noise and leakage current. Indeed, at a constant temperature, the dark current is also constant and cancels out in the differential ramp. This phenomenon is known as `persistence'. The measured signal represented in blue dots may thus be interpreted as the evolution of the persistence amplitude. This amplitude decreases rapidly, from 16\,$e^-$ at the beginning of the ramp to about 2\,$e^-$ after 140 seconds of acquisition. However, reaching 0\,$e^-$ will take several more hours.

\item In the illuminated differential ramp of Fig.~\ref{fig:antipersistence}, trapping initially dominates, leading to a signal (blue dots) lower than expected, whereas the signal  should follow the non-linearity observed further along the ramp (red dashes). This is referred to as `anti-persistence'. 

A correction for non-linearity is consequently needed to decorrelate both effects and will be introduced in Sect. \ref{sc:antipers} to address this issue.
 Precisely, the theoretical non-linear response of the photodiode (red dashes) is typically described by a decreasing affine function. At the beginning of the ramp, up to about 50 seconds, a signal below this affine function is observed. It is assumed that this lower signal reveals the presence of an anti-persistence phenomenon. We suppose that this phenomenon is more visible at the beginning of the ramp because its amplitude is higher then, similarly to the persistence signal. If charge trapping will definitely continue beyond this initial phase,  the non-linear photon measurement will eventually dominate, making anti-persistence more difficult to decorrelate. A correction for non-linearity is clearly needed to decorrelate both effects and will be introduced in Sect. \ref{sc:antipers}.

\end{itemize}

\subsubsection{\label{sc:Stateoftheart}Environmental dependencies and other behaviours}

Persistence amplitude and current exhibit dependencies on various environmental parameters, as documented in the literature, that we shall recall hereafter.

\paragraph{Prior fluence}
To begin with, let us examine the dependencies related to the previously received fluence. Based on the literature, several regimes for the persistence amplitude response have been identified depending on the fluence previously received by the MCT detectors.
\begin{itemize}
\item Case 1 (two regimes): following low fluences, the persistence response follows a linearly increasing regime before reaching a saturation plateau beyond a certain high fluence \citep{Tulloch2019,Baril2008}.
\item Case 2 (three regimes): Persistence exhibits 3 regimes: a first linearly increasing regime following low fluences, then a significantly steeper linear regime leading up to a saturation plateau \citep{Long2013,Serra2015, Serra2016}.  
\end{itemize}
In both cases, in the ``low fluence'' linear regime, persistence is proportional to the prior fluence received by the detector. 
The WIRCam team \cite{Baril2008} also compared the saturation regime for both cases: in case 2, the plateau is 10 times higher than in case 1. The adopted hypothesis posits a heterogeneous trap density, which could account for the significant increase preceding saturation. Conversely, the gradual increase of the persistence amplitude up to saturation (case 1) assumes a homogeneous trap density in the sensitive layer.

For the ALFA detector, given that it is a unique prototype, the persistence aspect has not received much research. Only the first case is observed in the literature \cite{LeGoff2022}.

\paragraph{Temperature} Regarding the temperature dependency, the ESO team working on the H4RG detectors for the MOONS instrument \cite{Ives2020} distinguish three different behaviours based on temperature. Below 50\,K, traps are frozen, resulting in very slow and weak persistence. Above 75\,K, thermal energy is sufficient to rapidly de-trap charges, reducing persistence. Between 50\,K and 75\,K, persistence becomes particularly challenging for astronomical applications. The thermal energy is sufficient to allow a partial release of trapped charges, but not enough to evacuate them quickly. This leads to a more pronounced persistence, especially for observations requiring high precision.
In general, the persistence increases with temperature \citep{Tulloch2018}. 
However, while higher temperatures accelerate de-trapping, they also increase the dark current \citep{Mosby2020}.

The chosen operating temperature for the \textit{Euclid} and SVOM mission detectors is around 100\,K. This choice is based, among other criteria, on a compromise between the reduction of persistence and the control of dark current.
At 100\,K, the de-trapping of charges is fast enough to limit the persistence effects. Although the dark current increases with temperature, cooling to 100\,K keeps it at an acceptable level for most astronomical applications ($< 0.01\,e^-/{\rm s}$).

\paragraph{Wavelength} Extensive testing for wavelength dependency has been conducted by various research teams, with the literature agreeing that no such dependency has been observed \citep{Baril2008,Rousset2024}. Current models describing persistence behaviour do not take wavelength into account.

\paragraph{Spatial variations} Regarding spatial dependencies, 
a team from the Space Telescope Science Institute \cite{Anderson2014} provided a 3D map of the traps of an H2RG detector, showing that the density of traps can vary by more than a factor of 20 over the entire matrix. Some regions exhibiting almost no persistence, while others (for example, near hybridisation defects) show consistently high trapping concentrations. 

The homogeneities are related to manufacturing defects. Among these defects, one can find impurities of the material, crystalline defects, or non-uniform doping profiles during the growth or processing of the MCT layer. Studies have demonstrated that traps exhibit spatial clustering, particularly at the MCT/CMOS interface, where mechanical stress can introduce additional defects \cite{Tulloch2018, Anderson2014}. They are also found in surface regions, where trapping densities are often higher at the interface with the coating layer.

Knowing these spatial variations is essential for applications like \textit{Euclid}, where pixel-to-pixel uniformity is important for photometry and high-precision spectroscopy.
The \textit{Euclid} NISP detector team has also mapped the persistence on the 16 H2RG detectors of the \textit{Euclid} mission, revealing significant spatial inhomogeneities~\cite{Kubik2024}. These variations are mainly attributed to the non-uniform distribution of traps in the detector material. Some areas with higher persistence may be related to hybridisation defects, such as voids in the adhesive layer between the MCT sensitive layer and the CMOS readout circuit. These defects could create localised regions with higher trapping densities, leading to increased persistence effects.

Our characterisation methodology, presented hereafter, aims to validate the dependencies observed in the literature, and to quantify them in the case of the two detectors under study. It also aims to provide new insights, particularly regarding anti-persistence, which has received limited interest to date.
This will enable us to quantify these dependencies for our detectors, paving the way for the development of a comprehensive model.

\section{Characterisation methodology of (anti-) persistence}
\label{sc:methodo}

The infrared detectors of the \textit{Euclid} and SVOM missions have been ground-tested on the dedicated facilities of the CPPM. We thus have a set of data allowing us to extract the performance parameters of the detectors. We propose to use these data for the study of (anti-)persistence phenomena and we have thus developed a methodology to extract the amplitude and current parameters of these two effects. These parameters were previously defined in Sect. \ref{sc:Persistence}.

Our approach consists in two major stages: firstly, a preparation of the raw data and secondly the actual extraction of (anti-)persistence parameters. The preparation itself includes several steps that will be described below: a pre-processing step, the implementation of the Savitzky--Golay filter to improve the signal-to-noise ratio (SNR) and the evaluation of environmental parameters (temperature, fluence...).
We will also describe how we measure the amplitude and current of persistence and anti-persistence. These extracted data will allow us to look into their behaviour as a function of time, temperature, flux or fluence, and wavelength. Preliminary results are presented in Sect.~\ref{sc:results}.

\subsection{Data optimisation and pre-processing}

Several pre-processing steps are applied to the raw data.
To mitigate the lack of data, as mentioned above, we have chosen to work with superpixels obtained by averaging 64 × 64 pixels. Indeed, since the illumination is in flat field, it is assumed that the pixels constituting a given superpixel have the same response. This grouping significantly improves the SNR.

Noise is further reduced thanks to an optimised method using reference pixels. This correction has been adapted to the specific configurations of the H2RG and ALFA detectors \citep{NouvelDeLaFleche2022} and typically reduces the readout noise (Sect. \ref{sc:Detection}).

Dead and hot pixels, identified as outliers, are masked according to characterisation maps established during tests. 

A final pre-treatment step is the conversion of ADU to electrons. This allows comparisons between detectors with different gain values, and expresses data into interpretable physical units. For this conversion, gain maps produced by Jean Le Graët are applied to the pre-processed data \citep{LeGraet2024}.

\subsection{Savitzky--Golay filtering}

Although the pre-processing significantly improves the quality of the data, the SNR of the ramp flux remains insufficient to reliably extract anti-persistence.
To improve the SNR, we further apply the Savitzky--Golay filter, which smooths the data while preserving its essential characteristics. This filter is based on a local polynomial regression, which makes it particularly efficient and adapted for our analysis. For each frame of a ramp, a local polynomial fit of the chosen degree is applied to the selected frames within the window. Then, the result is applied to the given frame.
The filtering thus requires the optimisation of two parameters: the degree of the polynomial and the window size.  In order to determine the optimal parameters, we compared the root mean square (RMS) errors for different window sizes and polynomial orders. Selecting the optimal parameters required finding a balance: minimising enough the RMS to smooth the curve, without over-fitting it to the point of reproducing noise.

\begin{figure}[!htbp]
  \centering
    \includegraphics[height=0.35\linewidth]{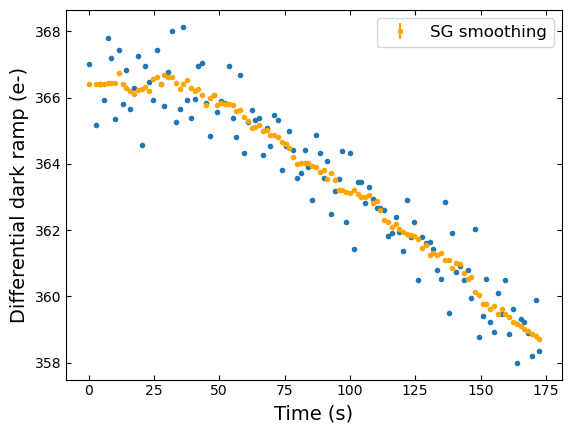}
    \vspace{0.1cm}
  \caption{H2RG detector: Comparison between raw and smoothed with Savitzky--Golay filter flux differential ramps. Set-point flux = 280\,$e^-/{\rm s}$, \textit{T} = $90$\,K. Smoothing parameters: window length of 13, polynomial order of 2} 
  \label{fig:smoothing}
\end{figure}

Figure \ref{fig:smoothing} gives an example of optimised smoothed ramp compared to its raw (not smoothed) differential ramp for a set-point flux of 280 $e^-/{\rm s}$ at 90\,K. It may be seen that smoothing effectively preserves the overall trend of the curve and highlights anti-persistence while reducing noise. 

A nuance appears at both ends of the ramp, where the window is incomplete. In these regions, the Savitzky--Golay filter still applies a local polynomial fit, but instead of using only the window centre point, the entire polynomial is preserved to ensure continuity. This processing allows not only to smooth the curve -- even at the ends -- but also to use this polynomial to extrapolate the curve if necessary. This extrapolation is particularly useful for analysing the beginning of the ramps, in order to study the (anti-)persistence at the first moments of the acquisition.

Finally, here is how we estimate the errors associated with this method.
We take into account the readout noise and the Poisson noise on each frame of the ramps. Thanks to the use of superpixels and the averaging of ramps acquired under identical conditions, these errors are mitigated.
To account for errors during smoothing\footnote{Let us note that though the filter is available in python3.xx libraries, it does not include the weighted error calculation, nor the extrapolation in zero.}, we have recoded an error-weighted Savitzky--Golay filter function, based on several articles on the subject \citep{Oxby2021a,Oxby2021b}. 

We have also added two other options: the calculation of the derivative and an extrapolation to zero.

For our example of Fig. \ref{fig:smoothing},  we have compared SNRs before and after the application of the Savitzky--Golay filter. This is calculated by dividing the signal power by the noise power. The SNR of the raw data (without smoothing)  is equal to 54.87\,dB, while that after smoothing is equal to 61.94\,dB. Applying the filter clearly improves the SNR.

Given all the advantages offered by this method and the SNR obtained after smoothing, it will be applied to all data used here.

\subsection{\label{sc:flux}Measurement of dependency parameters}

In this section we explain how we have measured the various parameters of interest to study the dependencies of (anti-)persistence, namely: detector temperature, wavelength, fluence, and flux of incident photons.

The temperature of the detector under test has been measured thanks to a calibrated Pt100 sensor located close to the detector. Moreover, the detector thermal stability is maintained to better than 1\,mK over more than 45 days thanks to an adequate design of the focal plane. 
The illumination of the detector is obtained from LEDs, whose central wavelength is provided by the manufacturer. Typically, the central wavelength of our LEDs lies within either the \HE or \JE bands. The cryostat itself has been designed to obtain an homogeneity better than 1\% over the whole detector \citep{Secroun2016}.

The measurement of an ``absolute'' incoming flux, or equivalently fluence, represents a greater challenge. It is particularly critical for the extraction of anti-persistence (Sect. \ref{sc:antipers}). Flux estimation would usually require a calibrated reference such as a photodiode, which was not available during the tests. Nevertheless we  used a non-calibrated reference photodiode for monitoring that does not allow absolute flux estimation \citep{Secroun2018}. Consequently, we chose to get an estimate directly from the ramps, paying particular attention to the selection of data.

\begin{figure}[!htbp] 
  \centering
  \includegraphics[height=0.35\linewidth]{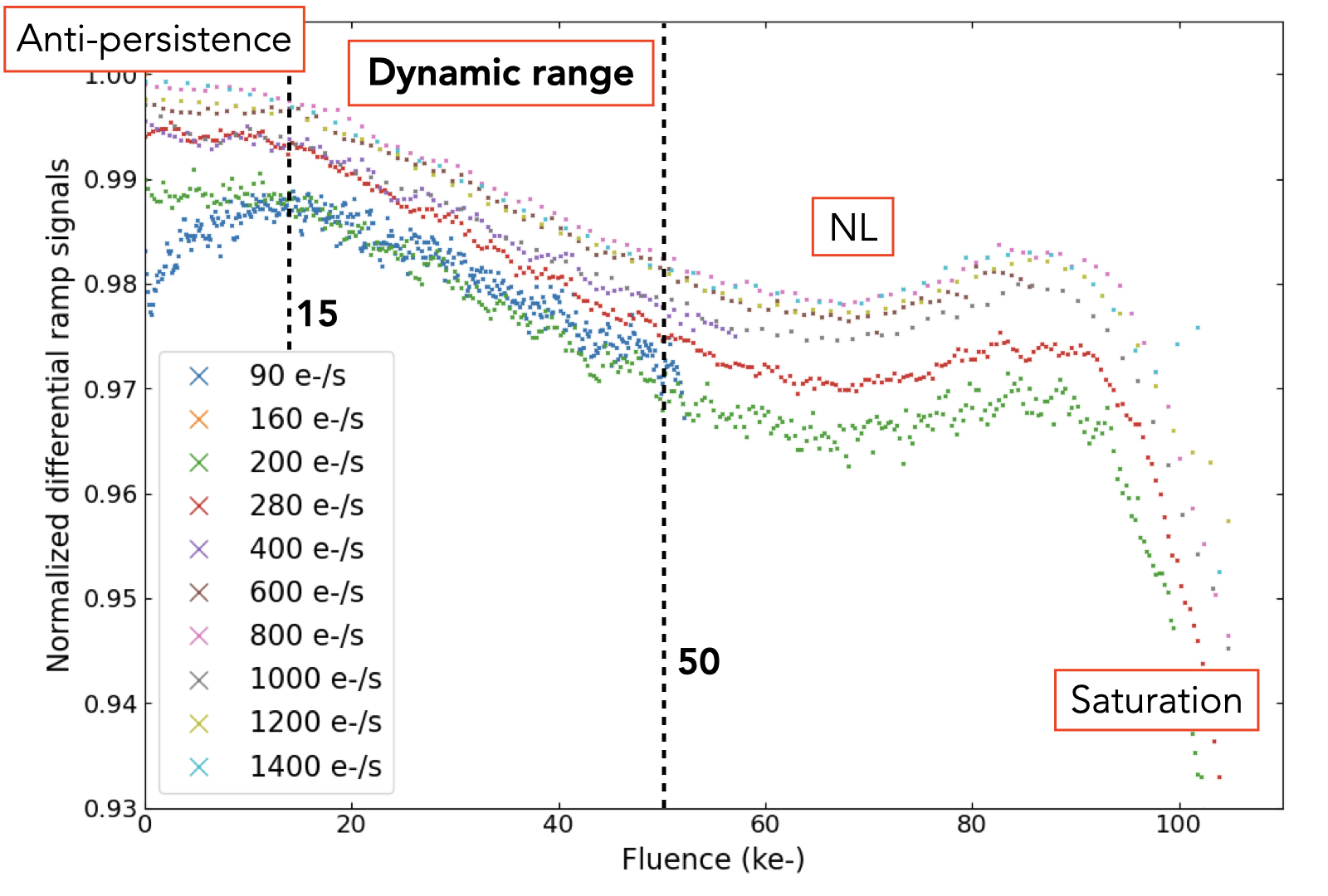} 
  \vspace{0.1cm}
  \caption{Normalised differential ramps of a flight H2RG detector,  at $T = 90$\,K. The different signatures of the detector's response (NL, anti-persistence, saturation) are identified on the graph.}
  \label{fig:responses}
\end{figure}

As may be seen in Fig.~\ref{fig:responses}, which shows various normalised differential ramps plotted as a function of fluence for a temperature of 90\,K, several signatures of the detector response may be recognised: 
\begin{itemize}
\item quadratic (second-order) non-linearity appears as a linearly decreasing line defining the response dynamic range, 
\item the signal at the end of the ramp drops to zero with saturation, because the signal no longer changes,
\item and at the very beginning of the ramp, a drop of the signal may be interpreted as anti-persistence.
\end{itemize}

For this specific detector, the response in the fluence range between 15 and 50\,k\,$e^-$ may be approximated locally by a quasi-linear fit.
Precisely, the limits of this range have been defined such that residuals remain below 1\%. The obtained range, defined as the detector dynamic range, mitigates effects from anti-persistence and higher order non-linearity regimes, and will used to estimate the incoming fluence. Its limits obviously need to be adapted to each detector.

At the end, the flux may be estimated by making a linear adjustment over the dynamic range, which is linearly decreasing. Its intercept should correspond to the measured flux lifted off the non-linearity.

\subsection{Extraction of (anti-)persistence parameters}
\label{sc:antipers}

To characterise anti-persistence and persistence, we respectively analyse differential flux ramps and differential dark ramps. Our aim is to extract the amplitudes in $e^-$ and currents in $e^-/{\rm s}$ associated with each phenomenon.

\subsubsection{Persistence}

As explained in Sect. \ref{sc:def}, in the case of persistence, the differential dark ramp directly gives the evolution of the amplitude of persistence. Indeed, no signal is expected in the absence of optical stimulation. Since the dark current is constant at the same temperature, its signal cancels in a differential ramp. It is therefore deduced that the unexpected signal observed in the differential ramp corresponds to persistence. The amplitude of persistence represents the number of de-trapped charges during a frame time. We suppose that the number of trapped charges is negligible in comparison. 

To determine the evolution of the persistence current, the derivative of this differential dark ramp is calculated using the Savitzky--Golay filter. Each point on the corresponding curve represents the evolution of the persistence current over time. This parameter represents the dynamics of de-trapped charges.

\subsubsection{Anti-persistence\label{sc:antipers}}

To study anti-persistence with minimal underestimation, a first treatment is necessary: the correction of non-linearity. 
As explained in the Sect. \ref{sc:def}, in differential flux ramps, non-linearity manifests as a decrease, while the anti-persistence observed at the beginning of these ramps follows an increasing trend. Therefore, the correlation between these two effects may lead to an underestimation of anti-persistence.

It should be noted that the correction of non-linearity will not allow a total decorrelation, since anti-persistence is present along the whole ramp. 

The proposed correction method, initially developed by the team from the Space Telescope Science Institute for the \textit{Hubble} Space Telescope \cite{Hilbert2014}, is based on a decomposition of the signal into orthogonal polynomials. Specifically, third-degree polynomials are used, as studies conducted by the \textit{Euclid} team have shown that higher-degree terms do not significantly alter the corrected results \cite{Euclid2026}.

Initial processing was performed to validate the method on H2RG data. Optimisation of this processing is ongoing; consequently, the full correction has not yet been applied to the entire data set.

This processing is applied within the previously defined dynamic range (between 15 k and 50 k$e^-$) to avoid treating anti-persistence as a non-linearity requiring correction. Once the process is applied, we get the coefficients of correction.
These coefficients are then applied to the entire ramp to amplify the anti-persistence signal, partially decorrelated from non-linearity.

After the correction, we have to extract the anti-persistence from the signal. To do so, the full ramp must be subtracted from a reference ramp, obtained using the flux measurement parameters (from the linear fit) estimated in the previous Sect. \ref{sc:flux}. Since the non-linearity correction has not yet been finalised, we subtract the full ramp from the reference ramp in the non-linear data (typically, the dashed red curve shown in Fig. \ref{fig:antipersistence} represents the reference ramp, and the blue signal represents the ramp).

To derive the anti-persistence current, we use the same method as for persistence by taking the derivative of the resulting subtracted ramp using the Savitzky--Golay filter. This current corresponds to the dynamics of charge trapping at the beginning of the acquisition.

According to the literature, persistence is most pronounced at the very beginning of dark acquisitions \citep{Serra2016, LeGoff2022}. To capture this behaviour, we extrapolate the amplitude and current curves to time zero using the Savitzky--Golay filter.

This section allowed us to measure the flux, apply an adapted pre-processing (including the Savitzky--Golay filter and the non-linearity correction), then extract the persistence and (anti-)persistence parameters. These steps have resulted in the creation of maps of (anti-)persistence parameters as well as associated error maps.

\section{Preliminary results}
\label{sc:results}
Now that the method has been detailed, we will look into some preliminary results to highlight the various dependencies of (anti-)persistence parameters.
We will focus on time $t = 0$, namely the very beginning of the acquisition, where (anti-)persistence is strongest.
The following subsections will present  the preliminary results organised by dependency: spatial response, fluence or flux, temperature, and wavelength. For each dependency, both persistence and anti-persistence will be addressed to highlight potential symmetries or differences in their behaviour.

\subsection{\label{tools}Tools and data}
Let us first describe the data to which we applied our methodology.
To measure (anti-)persistence, we need alternating ramps between illumination ramps and dark ramps. As said previously, this alternation allows us to isolate the two phenomena: persistence is observed in dark ramps following illumination, while anti-persistence is extracted from the differential flux ramps.

Data have been acquired with two types of illumination.
In the first case, a continuous and homogeneous illumination typically was obtained from a current-controlled LED. Here, acquisitions consisted of alternating illumination and dark ramps, where the length of each ramp has been varied. Illumination ramps reproduced typical \textit{Euclid} spectrometry ramps of 580 seconds while dark ramps were either 800 (so called `flux-dark' acquisitions) or 1160 frames long (so called `latency' acquisitions).

In the second case, the illumination was reduced to a short flash (shorter than a frame duration) at the beginning of the ramp. Then, we continue the acquisition without optical stimulation, and without reset: this is called soak. During this soak, we observe a decrease in the UTR ramp, even though it is supposed to be cumulative. It is assumed that this is due to delayed charge trapping. After a certain time, which depends on the flash fluence, the ramp increases again. It is then assumed that dark current and persistence dominate trapping. In general, the time to reach this balance is very long (several thousand seconds).

These two types of illumination are expected to modify the (anti-)persistence response.

To make sure that the illumination in flat field is uniform, the experimental setup is equipped with an integrated sphere, ensuring uniformity and stability of illumination.

The methodology that we described in Sect.~\ref{sc:methodo} has been applied to all dark ramps for persistence and to continuous illumination ramps for anti-persistence. Flash ramps shall require a dedicated processing function.

Table \ref{tab:detector_conditions} summarises the data used for the present study and the corresponding experimental conditions of acquisition, including  detector temperature range, central wavelength of illumination, illumination mode (continuous or flash), and fluence range.

\begin{table}[!htbp]
\centering
\caption{Experimental conditions during testing of H2RG and ALFA detectors.}
\label{tab:detector_conditions}

\begin{tabular}{@{} *{5}{>{\centering\arraybackslash}m{0.18\linewidth}} @{}}
\toprule
\textbf{Detector} & \textbf{Temperature} \textbf{(}K\textbf{)} & \textbf{Wavelength ($\mu m$)} & \textbf{Illumination mode} & \textbf{Fluence ($ke^{-}$)} \\
\midrule
H2RG & 80--90 & 1.55 (\HE band) & Continuous & 4.5--230 \\
\midrule
\multirow{2}{*}{ALFA} & 80--100 & 1.3 (\JE band), 1.65 (\HE band) & Continuous & 32--655 \\
     & 80--130 & 1.3 (\JE band) & Flash & 17--175 \\
\bottomrule
\end{tabular}

\end{table}

\subsection{Spatial response}

Let us first examine the (anti-)persistence spatial response
of both detectors. Figures \ref{fig:mapsh2rg}, \ref{fig:mapsalfa}, and \ref{fig:mapsalfaflash} show persistence amplitude maps for the two detectors considered here. We chose to show responses to a high prior fluence, a scenario where persistence effects are most pronounced and thus easiest to observe. These maps highlight the variation in persistence across the detectors arrays, where several effects may be visible. Hereafter we detail our observation for each detector separately.

\subsubsection{H2RG detector}

\begin{figure}[!htbp]
  \centering
  \includegraphics[height=5cm]{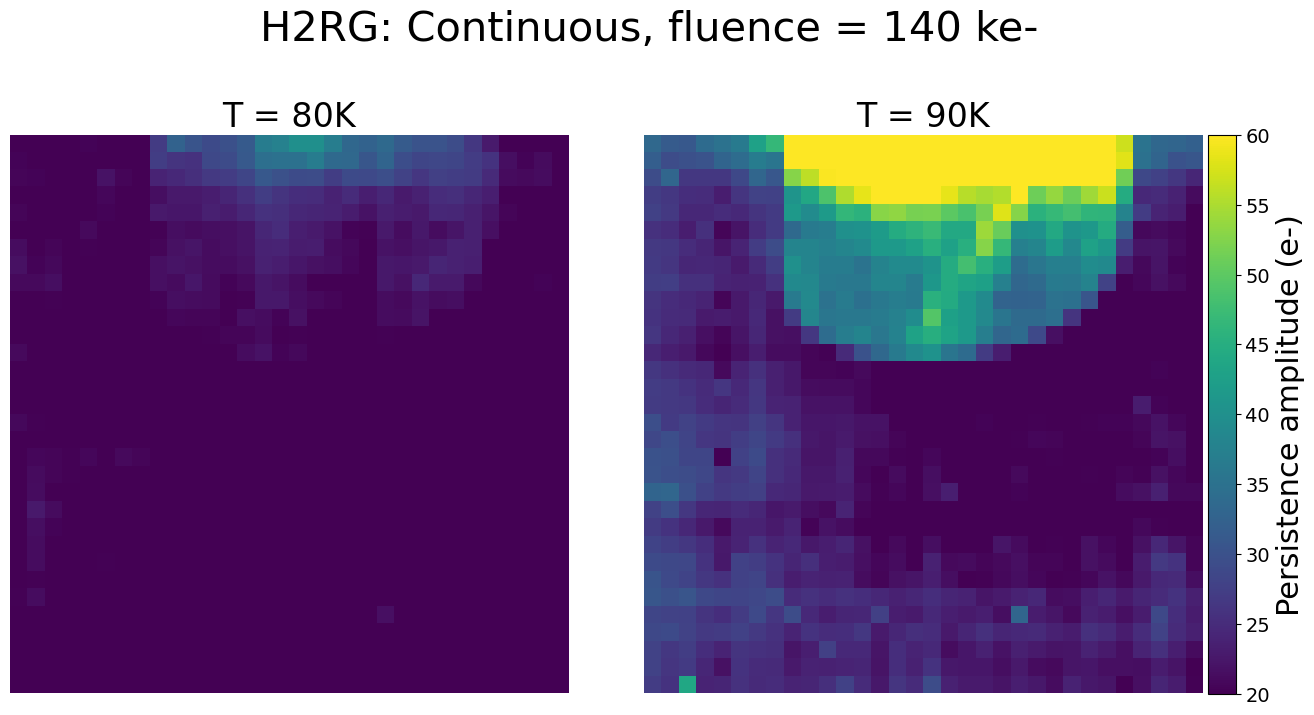}
  \vspace{0.1cm}
  \caption{H2RG's persistence amplitude following a continuous illumination of 140\,k$e^-$ fluence, \textit{T} = $ 80$ and $90$\,K, in \HE band. Maps are calculated over 64$\times$64 superpixels.}
  \label{fig:mapsh2rg}
\end{figure}

Figure \ref{fig:mapsh2rg} shows the persistence amplitude of our H2RG detector following a continuous illumination of 140\,k$e^-$ fluence at 80 and 90\,K, in the \HE band.  A spatial inhomogeneity is clearly visible in the northern part of the array, appearing to form an arc. This zone exhibits a higher persistence amplitude than the rest of the array, which is globally uniform. This inhomogeneity is observed both for the amplitudes and currents of persistence and anti-persistence. The possible causes of these inhomogeneities were discussed in the Sect. \ref{sc:Stateoftheart}.

These inhomogeneities mirror those described in the state-of-the-art literature \cite{Kubik2024}, where detector 18\,268 from the NISP instrument exhibits the same arc-like shape.

It can also be observed that persistence amplitude values increase with rising temperature, particularly within the inhomogeneous zone. 
Precisely, an increase of 10\,K in the homogeneous zone induces a persistence amplitude almost twice as high. 
The same temperature increase in the arc zone translates into a persistence amplitude multiplied by approximately 3.
The northern arc-shaped region may have a higher trap density, consistent with the observed spatial inhomogeneities in persistence amplitude.

\subsubsection{ALFA detector}

\begin{figure}[!htbp]
  \centering
    \begin{subfigure}{0.5\textwidth} 
    \centering
    \includegraphics[height=5cm]{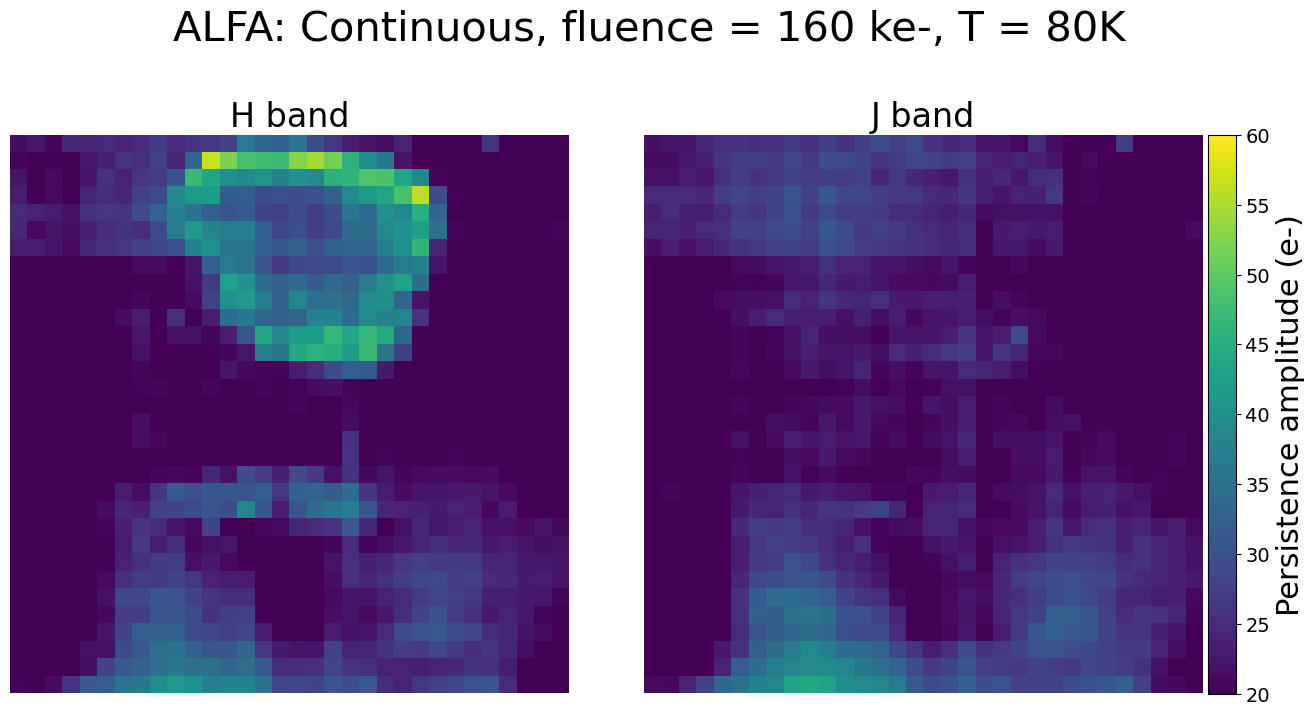}
     \label{fig:alfa80}
  \end{subfigure}
  \hfill
  \begin{subfigure}{0.5\textwidth}
    \centering
    \includegraphics[height=5cm]{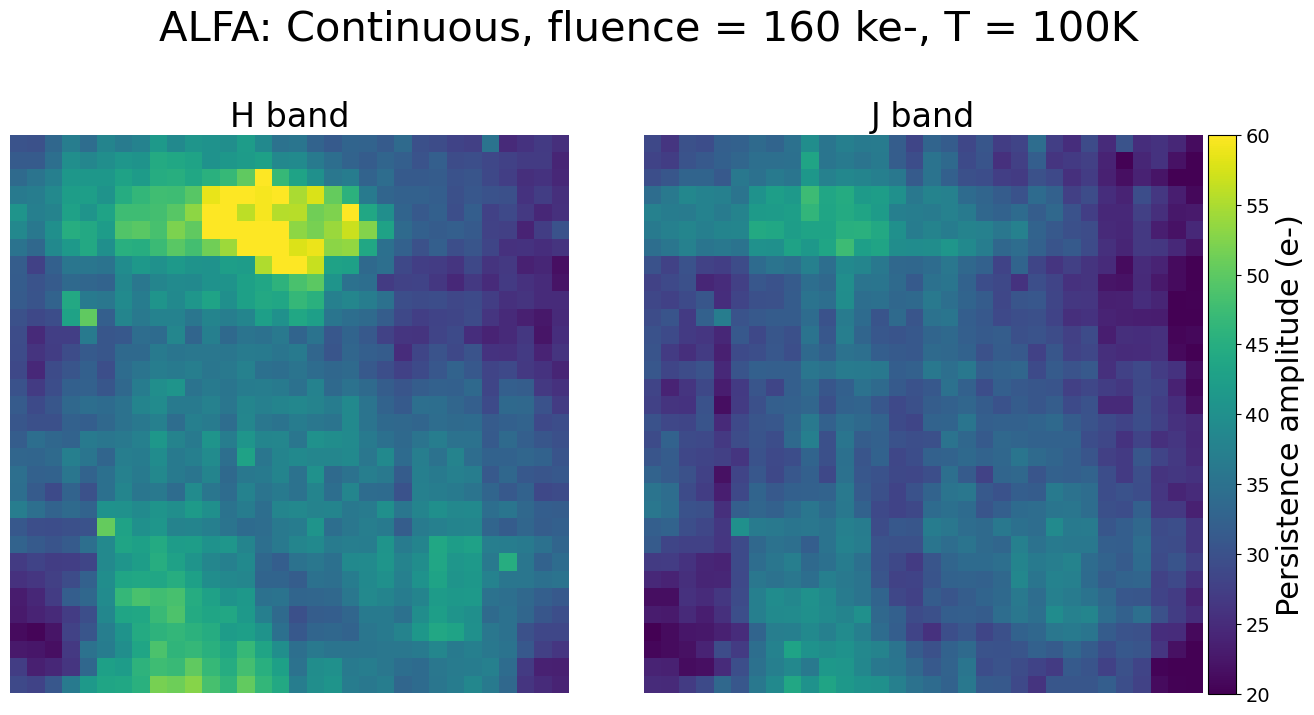}
    \label{fig:alfa100}
  \end{subfigure}
  \caption{ALFA's persistence amplitude following a continuous illumination of 160\,k$e^-$ fluence, at 80 and 100\,K and in \JE and \HE bands. Maps are calculated over 64$\times$64 superpixels.}
  \label{fig:mapsalfa}
\end{figure}

ALFA's persistence amplitude is shown in Fig. \ref{fig:mapsalfa} for various conditions of acquisition: temperatures of 80 and 100\,K and central wavelength in \JE and \HE bands. It may be seen that an increase in temperature, from 80 to 100\,K, not only causes the persistence amplitude to increase but also a change in the distribution of higher amplitude pixels, appearing in the form of a ring that seems to contract at 100\,K. An unexpected response is that to the wavelength of the illumination: clearly here new features appear in the \HE band where an annular inhomogeneity is observed compared with the response in the \JE band which is globally much flatter. This phenomenon is unexpected, since the literature mentions no wavelength dependency of persistence. The spectral dependence of persistence, although unexpected, can be attributed to the activation of traps as a function of their energy. In solid-state detectors, traps often have distinct activation energies, and photons of different wavelengths (corresponding to specific energies) can selectively excite charges from specific trap states. In this specific area of the detector, potential manufacturing defects appear to induce traps whose activation energy varies depending on the two tested wavelengths.

The same spatial inhomogeneities identified in the amplitude maps are also present in the current maps but are not shown here. These observations suggest that characteristic de-trapping times vary with these parameters. Indeed, the evolution of temperature may modify the activation energy of traps and the mobility of charges. The one of fluence may possibly probe traps located in different locations of the material, and changes in the wavelength may activate different traps, with different characteristic times of de-trapping. This could explain why certain effects, such as the ring observed, do not fully align with current theoretical expectations. If this were the case, we would not have observed any difference as a function of wavelength \cite{Rousset2024}.

\begin{figure}[!htbp]
    \centering
    \includegraphics[height=5cm]{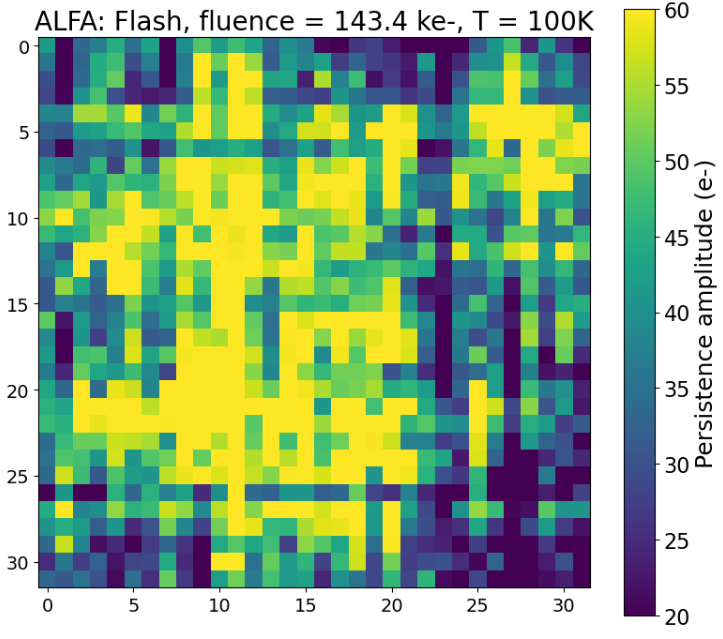}
    \vspace{0.1cm}
    \caption{ALFA's persistence amplitude following a flash illumination of 143\,k$e^-$ fluence, at  100\,K, in \JE band. Maps are calculated over 64$\times$64 superpixels.}
    \label{fig:mapsalfaflash}
\end{figure}

Figure \ref{fig:mapsalfaflash} shows a different pattern of persistence amplitude in response to a flash illumination. Overall, the persistence amplitude is higher than for continuous illumination.

Regarding anti-persistence, amplitude maps (not displayed here) do not reveal any specific feature like the ring seen in persistence maps.
This may be due to the fact that non-linearity is much more significant in the ALFA detector than in the H2RG.

Consequently, the anti-persistence signature is more difficult to capture. This can also mean that the anti-persistence is homogeneous on the array. This will be investigated after applying the non-linearity correction to all data.

\subsection{Analysis of environmental dependencies}

We now plot the persistence amplitude and current -- for many super-pixels -- as a function of fluence, using a colour code to indicate temperature and wavelength. This allows us to quantify these effects and gives us a base to model the behaviour of these dependencies. The knowledge of these dependencies will also help us design a new test plan to explore additional parameter values. 

The plots showing those dependencies are shown in Fig. \ref{fig:h2rg_pers} (H2RG persistence), \ref{fig:alfa_pers} (ALFA persistence), and \ref{fig:apers} (H2RG anti-persistence).

We selected the different areas discussed earlier for both detectors under study. This includes the ``ring'' region for the ALFA detector, that should give us insight into the dependency in wavelength (\JE and \HE bands), type of illumination (continuous or flash).

Similarly, for the H2RG, two areas were chosen: the homogeneous region (the lower half of the array) and the high-persistence region, allowing us to quantify and compare the persistence amplitude behaviour across these two areas.
As with the persistence maps, numerous observations can be made.

\subsubsection{Fluence dependencies}

For the study of fluence dependencies, we will exclusively analyse persistence. Rather, anti-persistence requires flux-based analysis since this phenomenon is only visible at the beginning of acquisition and will be presented in Sect. \ref{sc:flux}.

\begin{figure}[!htbp]
  \centering
  \begin{subfigure}{0.45\textwidth}
    \centering
    \includegraphics[height=0.7\linewidth]{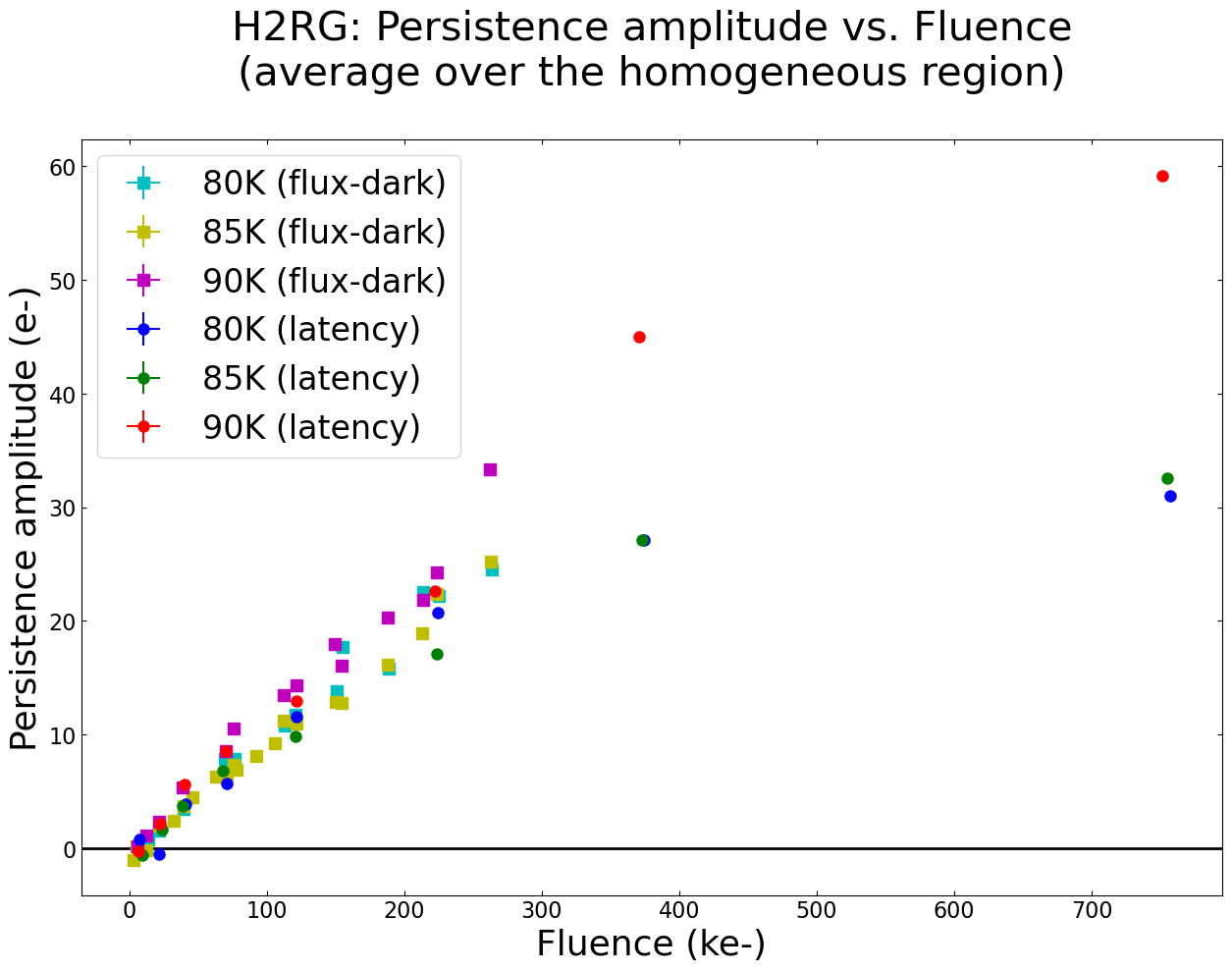}
    \caption{Low persistence homogeneous region}
    \label{fig:h2rg_pers_hom}
  \end{subfigure}
  \hfill
  \begin{subfigure}{0.45\textwidth}
    \centering
    \includegraphics[height=0.7\linewidth]{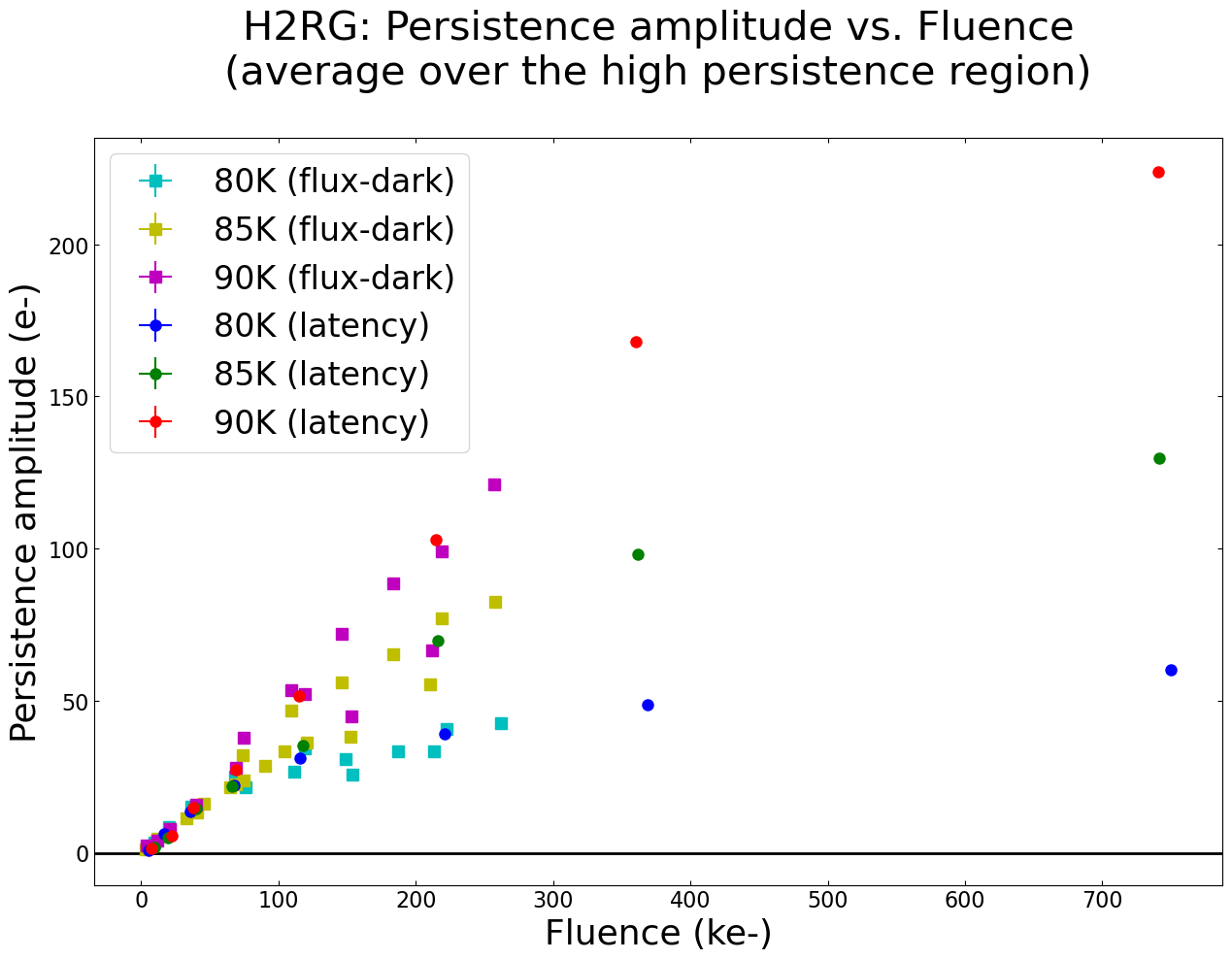}
    \caption{High persistence region}
    \label{fig:h2rg_pers_high}
  \end{subfigure}
  \vspace{0.4cm}
  \caption{ H2RG persistence amplitude vs. fluence at different temperatures (80, 85 and 90\,K) in \JE band. Two regions are compared: a large homogeneous region (4$\times$15 superpixels averaged) and the high persistence region (17$\times$31 superpixels averaged).}
    \label{fig:h2rg_pers}
\end{figure}

\paragraph{Persistence amplitude} 
For the H2RG detector, comparing the curve shapes on Figs. \ref{fig:h2rg_pers_hom} and \ref{fig:h2rg_pers_high} confirms that persistence amplitude values are indeed approximately three times higher in the high-persistence zone. 
It is also clear that towards low fluences, persistence follows a linear regime. At higher fluences, the curve shape distorts, suggesting the onset of saturation of traps with short characteristic time of de-trapping. This behaviour is observed starting at fluences of approximately 300\,k$e^-$ (ADC saturation). It should be noted that the signal for this detector saturates at 80\,k$e^-$. Unfortunately, we lack sufficient intermediate fluence values to observe this behaviour with good precision. The ratio between those two values of saturation seems close to 3, despite the limited data available at high fluences. 
These graphs also confirm that the heterogeneous zone is visible in all acquisitions and to verify that the persistence amplitude and current both depend on the previously received fluence.
We can also see that there is no particular difference between the flux-dark and latency data, except that the latency data allows us to explore higher fluences. The flux acquisition duration is constant for both types of data. Only the dark acquisition duration differs, and here we observe the persistence effect during the initial moments of the acquisition. Thus, no difference was expected.

\begin{figure}[!htbp]
  \centering
  \begin{subfigure}[t]{0.45\textwidth} 
    \centering
    \includegraphics[height=0.7\linewidth]{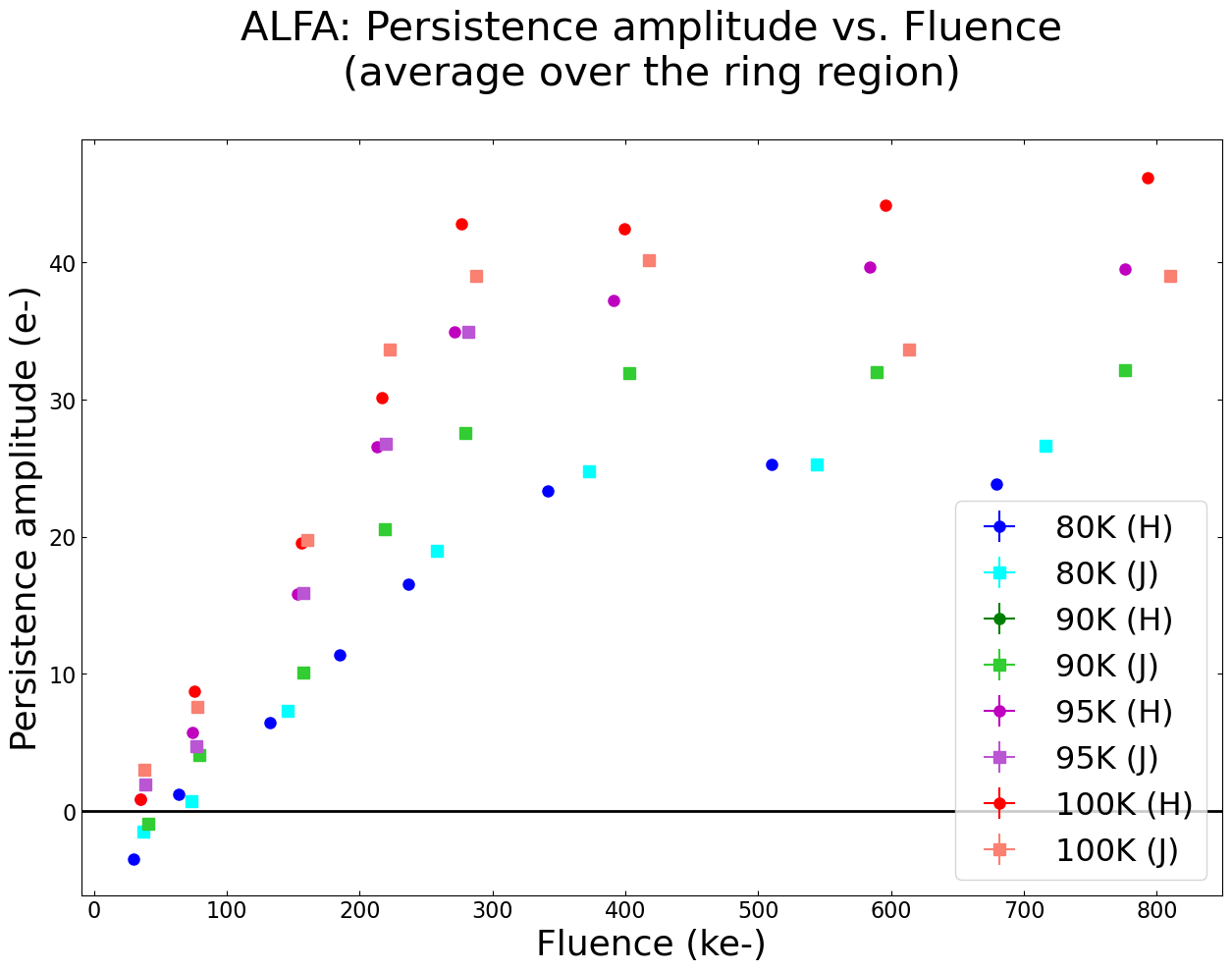}
    \caption{Ring region}
    \label{fig:alfa_pers_cont}
  \end{subfigure}
  \begin{subfigure}[t]{0.45\textwidth} 
    \centering
    \includegraphics[height=0.7\linewidth]{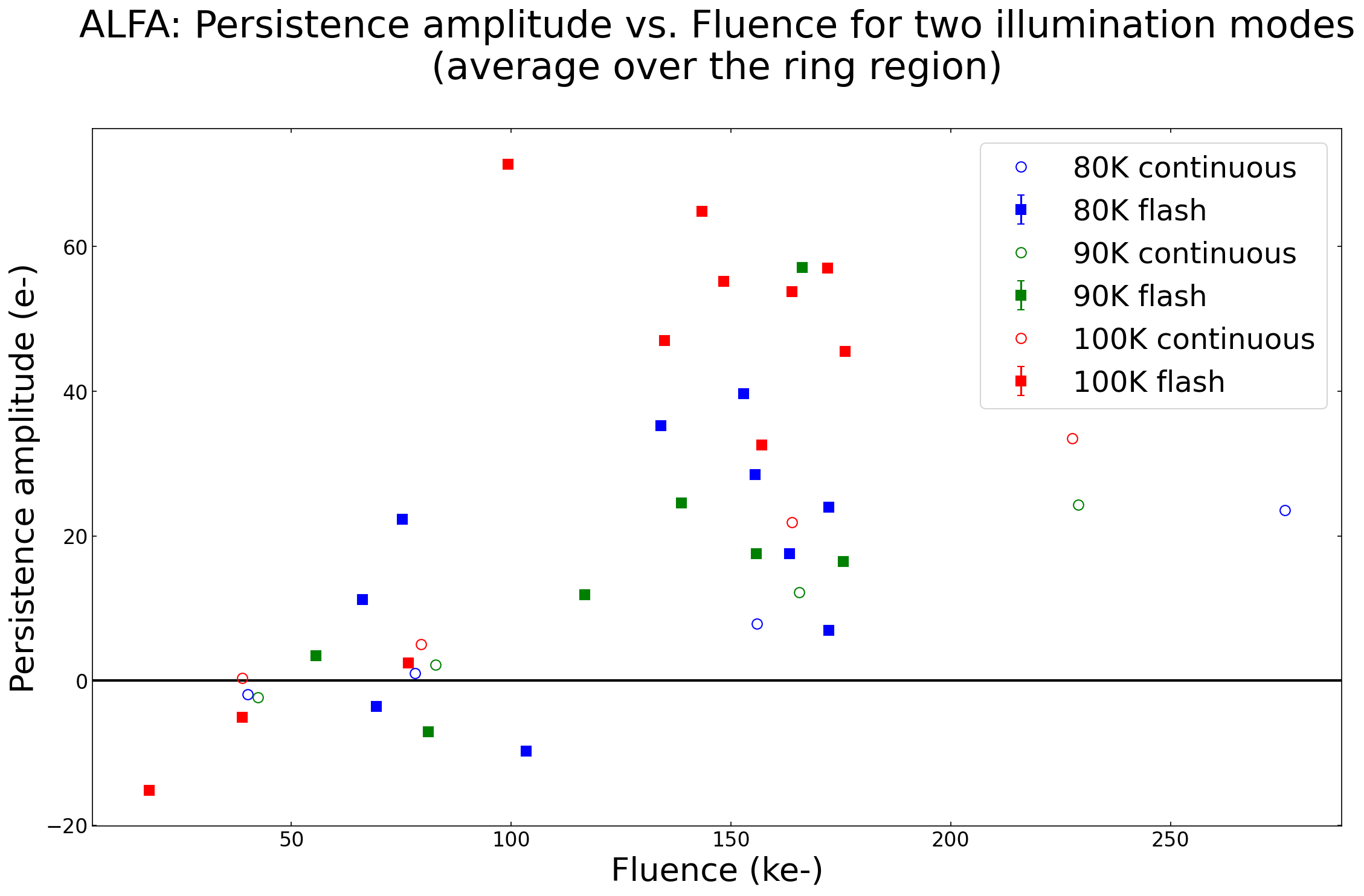}
    \caption{Ring region, linear range, \JE band}
    \label{fig:alfa_pers_flash}
  \end{subfigure}
  \vspace{0.4cm}

  \caption{ALFA persistence amplitude vs. fluence at different temperatures (80,90, 95 and 100\,K) in \JE and \HE bands. The graph on the right adds a comparison at low fluences of continuous (circles) and flash (squares) illuminations. Average over the ring region of 7$\times$3 superpixels.}
  \label{fig:alfa_pers}
\end{figure}

Regarding the ALFA detector, when looking at Fig. \ref{fig:alfa_pers}, we can observe a saturation of the persistence amplitude that becomes clearly visible at a fluence of approximately 300\,k$e^-$. The detector's ADC saturates at around 150\,k$e^-$; thus, the ratio between the persistence saturation level and the signal saturation level is 2. 

It can be noted that the limit value before saturation of 300\,k$e^-$ is the same as that observed in the H2RG detector data. We could have expected different values since the two detectors are different. One hypothesis is that there is a correlation between trap saturation and full-well capacity. For the H2RG, the full-well capacity is 130 k$e^-$, and for ALFA, it is 150 k$e^-$. In any case, the reason for this observation is still unknown, and deserves further study.

In the end, the persistence amplitude on the H2RG and ALFA detectors show similar behaviours: a linear regime is observed at low fluence, which saturates from 300\,k$e^-$.

On Fig. \ref{fig:alfa_pers_flash}, we can look at the effect of the illumination mode, flash or continuous. To make it easier to see the overall trend, linear fits are plotted. 

It can be seen that the points corresponding to the flash illumination are more scattered. This is due to the fact that we have fewer statistics on flash data. Indeed, in flash mode, we have only one ramp while in continuous mode, we average 20 ramps with the same acquisition conditions.

We can also observe a larger persistence amplitude following prior illumination in flash mode. For instance, at 100\,K and for a prior fluence of 160\,k$e^-$, the persistence amplitude is equal to 20 $e^-$ for a continuous illumination, and about 55 $e^-$ for a flash illumination.
Indeed, incorporating a very long soak time allows for the trapping of significantly more charges than continuous illumination lasting 145 or 580 seconds. Consequently, we expected to observe greater persistence.
Furthermore, by comparing the shapes of the linear fits, we can see that at 85\,K and 90\,K, the slope is steeper in the flash mode. Consequently, the variation in persistence amplitude as a function of fluence appears to depend on the illumination mode.

\paragraph{Persistence current}
The persistence current is consistent with that of the amplitude in both detectors: it increases with prior fluence and temperature, reaching a saturation threshold around 300\,k$e^-$.

\subsubsection{Flux dependencies}
\label{sc:flux}
\begin{figure}[!htbp]
  \centering

  \begin{subfigure}{0.45\textwidth}
    \centering
    \includegraphics[height=0.7\linewidth]{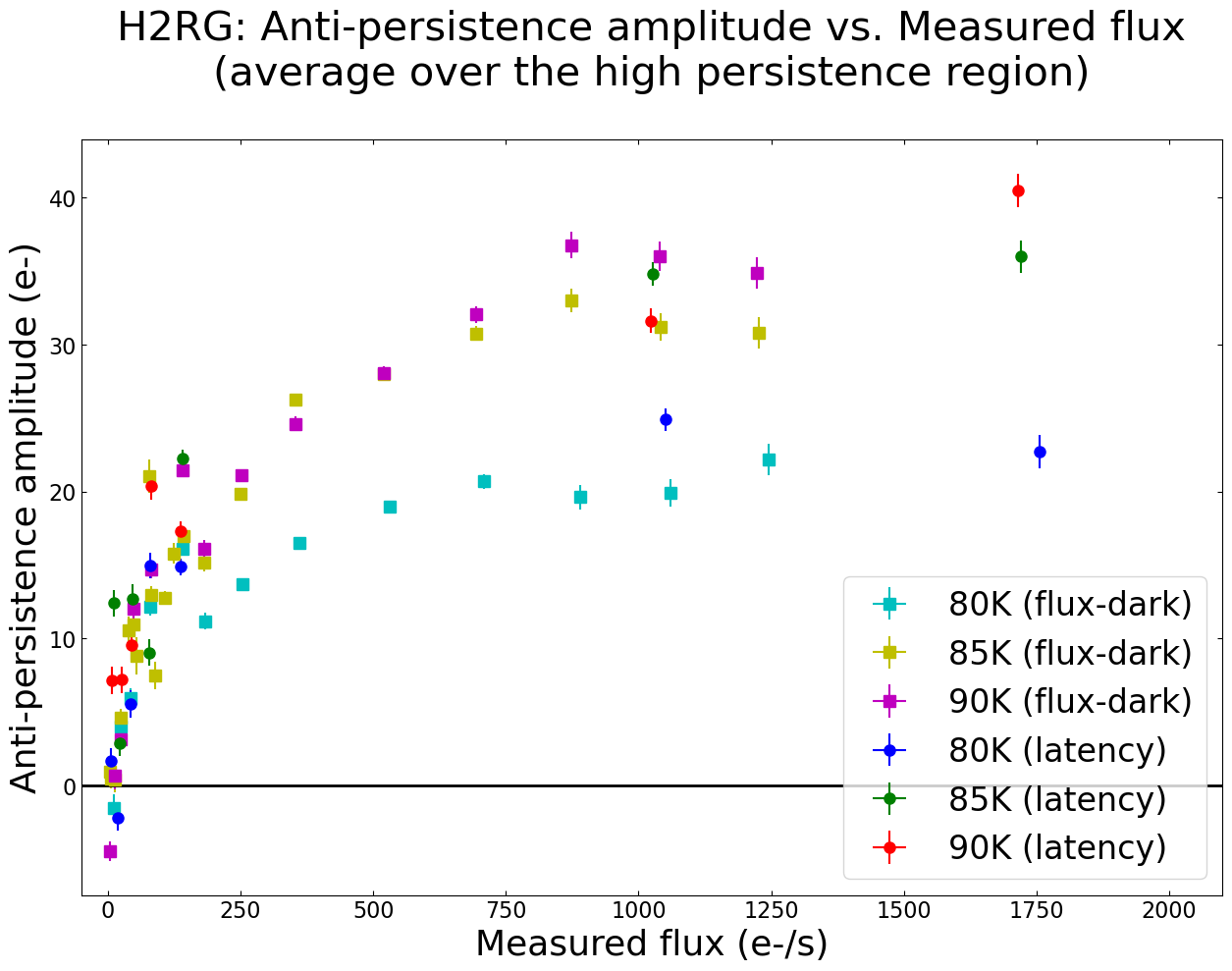}
    \caption{Anti-persistence amplitude}
    \label{fig:apers_amp}
  \end{subfigure}
  \hfill
  \begin{subfigure}{0.45\textwidth}
    \centering
    \includegraphics[height=0.7\linewidth]{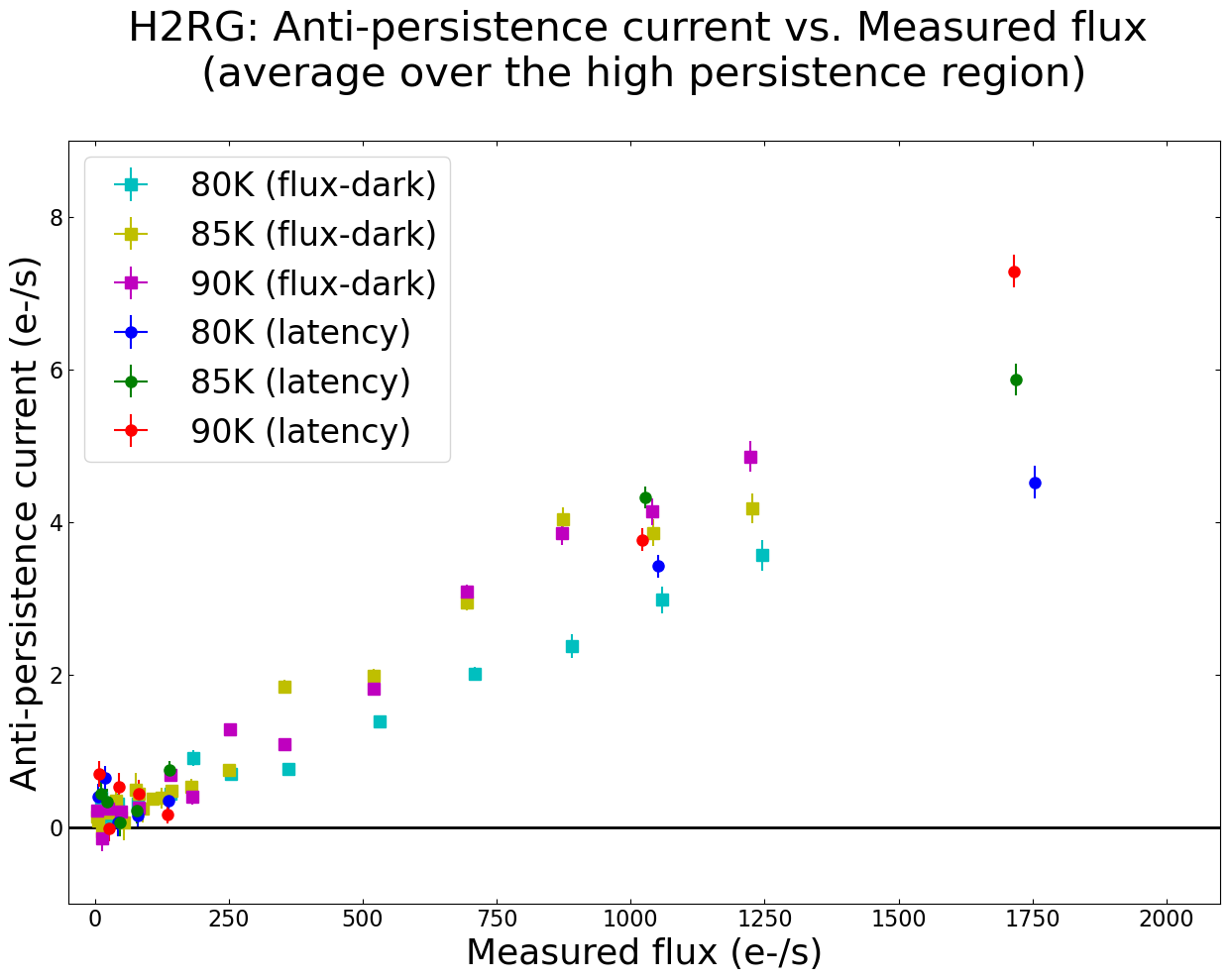}
    \caption{Anti-persistence current}
    \label{fig:apers_current}
  \end{subfigure}
  \vspace{0.4cm}
  \caption{H2RG anti-persistence amplitude and current: flux and temperature dependencies. 4$\times$15 superpixels over the high anti-persistence region of the detector are averaged.}
  \label{fig:apers}
\end{figure}

Let us now examine how anti-persistence depends on the measured flux. Fig.~\ref{fig:apers} shows two graphs illustrating these relationships: amplitude in Fig. \ref{fig:apers_amp} and current in Fig. \ref{fig:apers_current}. These results are from the H2RG detector, with the data averaged over the superpixels in the detector’s homogeneous region.

Looking at the graphs, both amplitude and current show an increasing trend as the flux increases:
At first glance, the anti-persistence amplitude increases sharply at low fluxes, then changes more slowly starting at around 500 $e^-/\rm s$. This suggests a logarithmic or exponential function.
One plausible explanation is that, at low flux levels, charge carriers are effectively trapped during the early stages of acquisition. As the flux increases, the characteristic short-time traps become saturated, leading to a stabilization of the amplitude.

The graph of current dependencies also shows interesting results. We observe a linear trend in this parameter as a function of flux, which suggests proportionality. This means that the charge trapping dynamics behave linearly with the flux.

In the case of the ALFA detector, anti-persistence data are more difficult to extract from the data. This is probably due to the fact that it exhibits higher order non-linearities than the H2RG detector. Since the results presented here do not include correction for non-linearity, anti-persistence is still underestimated.
In addition, the reading noise of the ALFA detector is higher: on average, it is 28 $e^-$, whereas for the H2RG detector, it is 13 $e^-$. The anti-persistence is therefore all the more difficult to extract from the ALFA detector data.

\subsubsection{Temperature and wavelength dependencies}

Figures \ref{fig:h2rg_pers}, \ref{fig:alfa_pers}, and \ref{fig:apers} may also give us some insight into the temperature and wavelength dependencies.
For both detectors, we observe what has already been seen in the analysis of spatial dependencies. Overall, the (anti-)persistence amplitude increases with temperature.

H2RG's temperature dependency is more pronounced in this highly persistent zone, compared to the homogeneous one, for both persistence and anti-persistence.
ALFA's temperature dependencies are also clearly visible. The same applies to wavelength dependency. Within the selected region, the \HE band exhibits a higher amplitude than the \JE band at 100\,K, though this distinction is less pronounced at lower temperatures. This may stem from the fact that the unit observed is electrons rather than photons. Indeed, QE depends on wavelength and could therefore affect the measured persistence amplitude in photons. We have not yet taken this into account, as we lack reliable QE measurements across the entire matrix.

The increase in the amplitude of persistence observed with temperature aligns with findings reported in the literature \cite{Tulloch2018,Ives2020}.
As thermal energy rises, phonon-assisted charge release becomes more efficient. Phonons -- vibrations of the crystal lattice -- provide the energy required for trapped charges to overcome the potential barriers of their trapping sites. This process accelerates the emission rate of trapped electrons, resulting in a stronger persistence signal at higher temperatures.

Regarding wavelength dependency, we can assume that switching to shorter (and therefore more energetic wavelengths) seems to preferentially activate certain traps located in the ring region. 

The proposed method for extracting (anti-)persistence parameters proves particularly effective, as it has revealed several key characteristics of the H2RG and ALFA detectors. Some of these results confirm behaviours already documented in the literature. We saw the increasing persistence parameters with temperature, and the different regimes depending on the fluence \cite{Baril2008}.

This method also shows unexpected phenomena, such as the wavelength dependency of the ring visible on the ALFA detector, which appear to contradict previous observations. These findings open up new avenues for refining our understanding of trapping mechanisms and, more broadly, persistence effects.

\section{Conclusions}

This study deals with a limitation on the IR detectors of the \textit{Euclid} and SVOM missions: the impact of (anti-) persistence. These phenomena deliver a signal that can last several hours, which distorts the flux measurements. 
We developed a characterisation methodology based on differential ramps, Savitzky--Golay filtering, and flux measurements to isolate and quantify these effects. This approach allows us to extract key parameters: amplitude and current. This extraction is done while decoupling from the other properties of the detectors, such as non-linearity.

Applied to ground-based datasets for the H2RG (\textit{Euclid}) and ALFA (SVOM) detectors, our method revealed significant spatial inhomogeneities. There are arc-shaped or ring-shaped regions of high persistence, probably related to manufacturing defects and non-uniform trap distributions. We also confirmed the environmental dependencies, including the saturation of persistence parameters at high fluence levels (higher than 300\,k$e^-$). These parameters are also affected by temperature, and unexpectedly by wavelength in ALFA, suggesting a wavelength-dependent activation of certain traps. The anti-persistence, although more difficult to extract, has been successfully isolated, exposing its exponential decay behaviour and correlation with non-linearity.

In the future, we will first finalize the method by applying non-linearity correction to all data on the H2RG and ALFA detectors. This should provide a better estimate of the anti-persistence parameters. Moreover, it would be interesting to redo this study with super-pixels smaller than 64x64, or even per single pixel. This would provide us with more accurate maps.
We will also extend the anti-persistence analysis to flash illumination tests, which promise to better elucidate its dynamics. Over the next year, these results will contribute to a physical model integrating all observed dependencies. This model will be useful for developing flight correction strategies, ensuring that \textit{Euclid} and SVOM achieve their scientific objectives with the required accuracy and reliability. 

\section*{Acknowledgements}

\AckEC

We express our gratitude to the reviewers from the \textit{Euclid} collaboration for their constructive comments on these proceedings.
We also extend our sincere thanks to our colleagues at the CPPM (Marseille Center for Particle Physics) for their collaboration and constant support throughout this study. Their feedback on this work has been extremely helpful.
This study was made possible thanks to CNES (National Center for Space Studies) and IN2P3 (National Institute of Nuclear and Particle Physics), which funded the current PhD thesis work.

\bibliography{bibliographiecorr}
\end{document}